\documentclass[conference,letterpaper]{IEEEtran}
\IEEEoverridecommandlockouts

\usepackage{amsmath,amssymb,amsfonts}
\usepackage{graphicx}

\usepackage{cite}

\usepackage{tikz}
\usetikzlibrary{positioning}
\usetikzlibrary{arrows.meta}
\usetikzlibrary{shapes.geometric}
\usepackage{algorithm}
\usepackage{algorithmic}

\usepackage{epstopdf} %converting to PDF
\usetikzlibrary{automata, positioning}
\newcommand{\RN}[1]{%
\textup{\uppercase\expandafter{\romannumeral#1}}
}
\begin{document}
\title{Chance-Constrained Joint Beam Hopping Scheduling and Bandwidth Allocation for LEO Satellites}
\IEEEoverridecommandlockouts 
% \author{Authors}

\author{
  \IEEEauthorblockN{Olfa Ben Yahia\IEEEauthorrefmark{1}, Jean-François Frigon\IEEEauthorrefmark{1}, Brunilde~Sansò\IEEEauthorrefmark{1}, Stéphane Martel\IEEEauthorrefmark{2}, Guillaume Mantelet\IEEEauthorrefmark{2}, \\ Antoine~Lesage-Landry\IEEEauthorrefmark{1}, and Gunes~Karabulut~Kurt\IEEEauthorrefmark{1}}

    \IEEEauthorblockA{\IEEEauthorrefmark{1}Poly-Grames Research Centre, Department of Electrical Engineering, Polytechnique Montréal, Montréal, QC, Canada\\
% \IEEEauthorrefmark{3}GERAD, Montréal (Qc), Canada, H3T 1J4\\
  \IEEEauthorrefmark{2}MDA, 21025 Trans-Canada Hwy Sainte-Anne-de-Bellevue, Qc H9X 3R2.\\
  \textit{\{olfa.ben-yahia, j-f.frigon, brunilde.sanso, antoine.lesage-landry, gunes.kurt\}@polymtl.ca,} \\  \textit{\{Stephane.Martel, Guillaume.Mantelet\}@mda.space 
  }}}
\maketitle
\begin{abstract}
Service level agreement (SLA) reliability is a key requirement in multibeam satellite systems, where operators must satisfy heterogeneous and time-varying traffic demands using limited resources. Beam hopping (BH) enables flexible resource allocation by activating only a subset of beams at each time slot. However, most existing BH optimization methods rely on predicted channel conditions and do not explicitly control the probability of SLA violation under spectral efficiency uncertainty. This work introduces an SLA-aware chance-constrained BH optimization model for joint beam scheduling and bandwidth allocation. The proposed formulation maximizes the expected delivered traffic while ensuring that each cell demand is satisfied with a target reliability level. By modelling the spectral efficiency prediction error as Gaussian, the
probabilistic SLA constraints are reformulated as deterministic second-order
cone constraints. Evaluation results demonstrate that the proposed method maintains the target SLA reliability under increasing traffic load and spectral efficiency uncertainty, while maintaining fairness and efficient resource
allocation.
\end{abstract}


%%**************************************Intro***********%%
\section{Introduction}

Service level agreements (SLAs) are becoming increasingly important for
satellite operators, as they are directly related to service reliability,
customer satisfaction, and revenue maximization. While traditional SLAs were
often defined as static agreements, future satellite networks require more
adaptive SLA mechanisms that can react to changing traffic demand, resource
availability, and service requirements \cite{bukhari2024dynamic}. To maximize
their revenues and avoid SLA violations, satellite operators must therefore
allocate the available radio resources efficiently, dynamically, and reliably \cite{yungaicela2026rslaq}.
This requirement is particularly critical in multibeam satellite systems, where
traffic demand is heterogeneous, unevenly distributed, and time-varying across
the coverage area. As a result, SLA management becomes challenging because the
operator must improve system performance while ensuring that service guarantees
are maintained under uncertain and dynamic operating conditions.

Given the high cost and limited availability of satellite resources, efficient
resource allocation and time sharing are essential in next-generation multibeam
satellite systems. These systems require flexible mechanisms to distribute
on-board power, bandwidth, and time across the coverage region to handle increasing traffic demand and the emergence of satellite-to-device services, which introduce irregular and time-varying traffic patterns with
diverse performance requirements. In this regard, programmable regenerative payloads have been identified as a key enabler of flexible onboard processing, dynamic resource management, and adaptive service provisioning in future high-throughput satellite systems \cite{10648786,10787026}. In addition, beam hopping (BH) has emerged as
a promising technique for adaptive resource management in multibeam satellite
systems. Unlike conventional fixed illumination schemes, where all beams are continuously activated with a fixed resource distribution, BH activates only a subset of beams during each time slot. The set of
active beams, their illumination duration, and the allocated resources can be adapted over time according to traffic demand, channel conditions, and system constraints. 
This time-domain flexibility enables the satellite payload to concentrate resources on high-demand areas while avoiding inefficient resource allocation to lightly loaded cells. It can also help mitigate co-channel interference, improve system throughput, and adapt to varying conditions across different scenarios \cite{abudureheman2025resource}.
\begin{figure*}[t]
  \centering  
  \includegraphics[height=0.28\textheight,keepaspectratio]{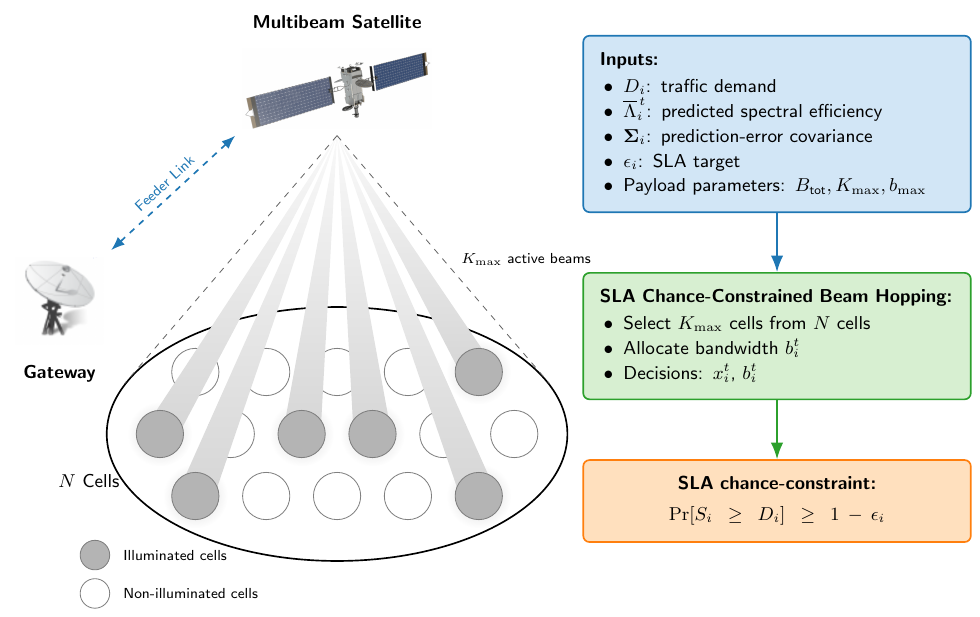}
  \caption{System model of the proposed SLA chance-constrained BH framework.}
  \label{fig:system_model}
\end{figure*}
The main challenge in BH is therefore to determine an efficient resource allocation policy, namely, which beams to activate, when to activate them, and how much resource to allocate to each active beam. In the current literature, BH resource allocation has been studied from different perspectives. Several works have investigated dynamic BH and resource allocation for non-geostationary orbit (NGSO) satellite
systems, including joint beam scheduling and power allocation, traffic-aware
resource management, load balancing, and bandwidth allocation
\cite{chen2024joint,jia2025dynamic,zheng2024traffic,wang2025resource,10570741}.
More recent studies have further considered demand-aware BH, distributed BH strategies, and joint illumination, power, and band allocation in low Earth orbit (LEO)/NGSO satellite networks
\cite{zhao2025demand,li2026distributed,zamacola2026joint,kim2026dynamic}.
These works demonstrate BH's potential to improve quality-of-service (QoS) metrics and key performance indicators (KPIs), including throughput, delay, spectral efficiency, load balancing, and resource utilization.

Despite these advances, most existing BH optimization frameworks assume that perfect or accurately predicted system information, such as channel state information (CSI) or spectral efficiency, is available when scheduling decisions are computed \cite{wang2023efficient,chen2024joint}. Although this assumption simplifies the optimization problem, it is not fully realistic in practical satellite networks, where the realized spectral efficiency may deviate from its predicted value due to channel variations, satellite motion, pointing errors, fading, and atmospheric effects. Consequently, deterministic allocation based only on predicted conditions may be insufficient for effective SLA management. Such uncertainty can be modelled by expressing the realized spectral efficiency as the sum of a predicted nominal value and an additive prediction error, often represented as a Gaussian random variable \cite{5164911}. Similar observations have motivated robust and probabilistic resource allocation approaches for satellite and satellite-terrestrial networks under channel uncertainty \cite{li2018robust}. In this work, an optimization under uncertainty approach is used to incorporate the available stochastic information into the resource-allocation problem~\cite{li2008chance}. In particular, chance-constrained programming imposes probabilistic constraints on users' QoS and SLA requirements, thereby controlling the probability of SLA violations under uncertain operating conditions. Thus, a BH solution that satisfies traffic demand under predicted conditions may still violate SLA requirements when the actual system conditions change, whereas a chance-constrained formulation makes it possible to quantify the trade-off between resource efficiency and service reliability.

%Therefore, deterministic BH optimization alone cannot guarantee that the probability of SLA violation remains below a predefined target. 

The high flexibility offered by BH has also attracted attention in the latest standardization and industrial developments. Recent standards such as 5G-NTN and DVB-S2X, as well as
industrial players, including MDA Space with its MDA AURORA\texttrademark{} digital beamforming architecture \cite{mda_aurora_2025}, are incorporating flexible beam management and BH-related capabilities into
next-generation satellite designs. This evolution further emphasizes the need for optimization models that can exploit BH flexibility while maintaining reliable service delivery.

Motivated by these limitations, this work proposes an SLA-aware chance-constrained BH resource allocation model. The objective is to exploit the flexibility of BH while explicitly controlling the probability of demand violation under spectral efficiency uncertainty.
The main contributions of this paper are summarized as follows:
\begin{itemize}
    \item We first formulate a baseline BH optimization problem, where beam illumination and bandwidth allocation are jointly optimized to maximize the delivered traffic under demand and resource constraints without explicitly accounting for spectral efficiency uncertainty.

    \item We then extend this formulation into an SLA-aware chance-constrained problem that accounts for spectral efficiency uncertainty and guarantees demand satisfaction with a target reliability level, thereby reducing the probability of SLA violation.

    \item We derive a deterministic reformulation of the SLA chance-constrained problem under a Gaussian prediction-error model, where a reliability margin is introduced to account for uncertainty.

\item We evaluate the proposed SLA-aware BH approach against the baseline model and offered-capacity-to-demand ratio (OCDR) \cite{9146871} through Monte Carlo simulations. The results demonstrate that the proposed method enhances service reliability by reducing the probability of SLA violations while preserving efficient resource utilization, limiting unmet demand, and ensuring fairness across cells.
\end{itemize}

The remainder of this paper is organized as follows. Section~\ref{SEct2}
presents the system and channel models. Section~\ref{SEct3} introduces the baseline BH optimization problem. Section~\ref{SEct4}
describes the proposed SLA-aware chance-constrained formulation. Numerical results and discussion are presented in Section~\ref{sect5}. Finally, Section~\ref{sec:conclusion} concludes the paper.

%%**************************************Model***********%%

\section{System and Channel Models}
\label{SEct2}
In this work, we consider downlink data communications in a regenerative NGSO multibeam
satellite system, as illustrated in Fig.~\ref{fig:system_model}. The satellite
serves a coverage area divided into multiple ground cells through a multibeam
payload. Due to payload and resource limitations, only a subset of beams can be
activated during each time slot.
The service footprint is partitioned into \(N\) ground cells indexed by
$\mathcal{I}=\{1,2,\dots,N\}$.
The BH cycle is organized over a scheduling horizon of \(T\)
consecutive time slots, each of duration \(T_{\mathrm{slot}}\). The set of time
slots is denoted by
$\mathcal{T}=\{1,2,\dots,T\}$.
We assume a fixed beam-to-cell association framework. Hence, the beam \(i\) is permanently associated with cell \(i\). Therefore, the scheduler decides at each
time slot whether the beam associated with the cell\(i\) is activated or not. Let \(x_i^t\in\{0,1\}\) denote the illumination variable, where
\(x_i^t=1\) if a cell \(i\) is illuminated during a time slot \(t\), and
\(x_i^t=0\) otherwise. Only a subset of beams, denoted by~\(K_{\max}\), can
be simultaneously activated at each time slot.
At the cell level, the large-scale channel power gain~$g_i^t$ between the satellite
beam and cell \(i\) during the time slot \(t\) is
expressed as
$g_i^t = G^i_{t,\mathrm{TX}}G^i_{t,\mathrm{RX}}L_i^t,$
where \(G^i_{t,\mathrm{TX}}\) and \(G^i_{t,\mathrm{RX}}\) denote the transmit
and receive antenna gains associated with cell \(i\), respectively, and
\(L_i^t\) is the large-scale path loss given by
  $ L_i^t =
\left(
\frac{c}{4\pi d_i^t f_\text{c}}
\right)^2,$
where \(d_i^t\) denotes the propagation distance between the satellite and
cell \(i\), \(f_\text{c}\) is the carrier frequency, and \(c\) indicates the speed of light.
We assume that the transmit power $P_{\mathrm{tot}}$ is fixed and equally shared among the maximum number of active beams. This assumption preserves a tractable linear problem structure, because joint bandwidth and power optimization would lead to a nonlinear formulation, which will be considered in future work. Accordingly, the transmit power per active beam is given by
$P_0=\frac{P_{\mathrm{tot}}}{K_{\max}}$ and the instantaneous signal-to-noise ratio (SNR) can be written as   $ \gamma_i^t =
\frac{P_0 g_i^t}{\sigma_i^2}$, where \(\sigma_i^2\) is the effective noise power associated with cell \(i\).
The corresponding spectral efficiency is given by
\begin{equation}
\Lambda_i^t =
\log_2\left(1+\gamma_i^t\right).
\label{eq:spectral_efficiency_fixed}
\end{equation}
Let \(b_i^t\in[0,1]\) denote the normalized bandwidth fraction allocated to
cell \(i\) during slot \(t\). The parameter \(b_{\max}\) denotes the maximum
bandwidth fraction that can be allocated to one active cell in a given slot, i.e., $b_i^t \leq b_{\max}, \forall i,t $.
The bandwidth assigned to cell \(i\) is then expressed as 
$  B_i^t = B_{\mathrm{tot}} b_i^t.$

%%%%%%%%%%%%%%%%%%%%%%%%%%%%%%%%%%%%
\section{Proposed Beam-Hopping Problem}
\label{SEct3}
The traffic demand distribution across the service footprint is an important
input for the design of BH patterns, as it indicates how the available
resources should be distributed among the cells over time. Under BH operation,
a cell is not necessarily illuminated in every time slot. Therefore, demand
satisfaction is imposed over the complete BH cycle rather than on a per-slot
basis. Let \(D_i\) denote the aggregate traffic demand of cell \(i\in\mathcal{I}\)
over the complete BH cycle. The scheduler must ensure that the accumulated
delivered traffic to each cell over the complete cycle satisfies this demand.
For a given spectral efficiency \(\Lambda_i^t\), the traffic delivered to cell
\(i\) during time slot \(t\) is
\(T_{\mathrm{slot}}B_{\mathrm{tot}}b_i^t\Lambda_i^t\). Therefore, the
accumulated delivered traffic over the complete BH cycle is obtained by
summing this quantity over all time slots.

Let
\(\mathbf{x}^t=[x_1^t,\ldots,x_N^t]^{\mathsf{T}}\)
and
\(\mathbf{b}^t=[b_1^t,\ldots,b_N^t]^{\mathsf{T}}\)
denote the illumination and bandwidth-allocation vectors at time slot \(t\).
The baseline deterministic BH problem is then formulated as
\begin{subequations}
\label{eq:P1_fixed}
\begin{align}
\max_{\substack{\mathbf{x}^t,\mathbf{b}^t\\ t\in\mathcal{T}}}\quad
& \sum_{t\in\mathcal{T}}\sum_{i\in\mathcal{I}}
T_{\mathrm{slot}} B_{\mathrm{tot}} b_i^t \Lambda_i^t
\label{eq:P1_fixed_obj}
\\
\text{s.t.}\quad
& \sum_{t\in\mathcal{T}}
T_{\mathrm{slot}} B_{\mathrm{tot}} b_i^t \Lambda_i^t
\ge D_i,
\qquad \forall i\in\mathcal{I},
\label{eq:P1_fixed_demand}
\\
& \mathbf{1}^{\mathsf{T}}\mathbf{x}^t \le K_{\max},
\qquad \forall t\in\mathcal{T},
\label{eq:P1_fixed_active_beams}
\\
& \mathbf{1}^{\mathsf{T}}\mathbf{b}^t \le 1,
\qquad \forall t\in\mathcal{T},
\label{eq:P1_fixed_bandwidth_budget}
\\
& \mathbf{0} \le \mathbf{b}^t \le b_{\max}\mathbf{x}^t,
\qquad \forall t\in\mathcal{T},
\label{eq:P1_fixed_bandwidth_activation}
\\
& \mathbf{x}^t\in\{0,1\}^{N},
\qquad \forall t\in\mathcal{T},
\label{eq:P1_fixed_binary}
\end{align}
\end{subequations}
where the objective function in \eqref{eq:P1_fixed_obj} maximizes the total delivered
traffic over all cells and time slots,
\eqref{eq:P1_fixed_demand} ensures that the aggregate demand of each cell is
satisfied over the complete BH cycle,
\eqref{eq:P1_fixed_active_beams} limits the number of simultaneously active
beams to \(K_{\max}\) in each time slot,
\eqref{eq:P1_fixed_bandwidth_budget} ensures that the total allocated
bandwidth fraction does not exceed the available system bandwidth, and \eqref{eq:P1_fixed_bandwidth_activation} is applied element-wise and
couples the illumination and bandwidth-allocation variables. Specifically, if \(x_i^t=0\), then \(b_i^t=0\), meaning that no bandwidth is assigned to a non-illuminated cell. If \(x_i^t=1\), then
the allocated bandwidth fraction is upper bounded by \(b_{\max}\), which
prevents one active cell from occupying an arbitrarily large portion of the
total bandwidth. Finally, \eqref{eq:P1_fixed_binary} defines the
binary illumination decision.
The given formulation assumes that the spectral efficiency values
\(\Lambda_i^t\) are perfectly known at the transmitter.
However, in practical NGSO
satellite systems, these values are obtained from channel prediction and may
deviate from the realized values due to prediction errors, satellite motion,
pointing variations, fading, and atmospheric effects. Therefore, a solution that
satisfies the demand under these predicted conditions may lead to
SLA violations. This motivates the SLA chance-constrained formulation introduced in the next section.
%%%%%%%%%%%%%%%%%%%%%%%%%%%%%%%%%%%%%%%%%%%%%%%%%%%%%%%%%%%%%

\section{SLA Chance-Constrained Formulation}
\label{SEct4}
To account for channel uncertainty, the spectral efficiency is represented as a
random vector in this section. Let
\(\boldsymbol{\Lambda}_i =
[\Lambda_i^1,\Lambda_i^2,\ldots,\Lambda_i^T]^{\mathsf{T}}\)
denote the random spectral efficiency vector of cell \(i\) over the complete
BH cycle, and let
\(\mathbf{b}_i=[b_i^1,b_i^2,\ldots,b_i^T]^{\mathsf{T}}\)
denote the corresponding bandwidth allocation vector.
We assume that the scheduler relies on a prediction model to estimate the
spectral efficiency over the BH cycle. The realized spectral efficiency vector is modelled as the sum of the predicted
spectral efficiency vector and a prediction error, i.e.,
\begin{equation}
\boldsymbol{\Lambda}_i =
\overline{\boldsymbol{\Lambda}}_i
+
\mathbf{e}_i.
\label{eq:spectral_prediction_error}
\end{equation}
%\vspace*{0.04in}

In (\ref{eq:spectral_prediction_error}), \(\overline{\boldsymbol{\Lambda}}_i\) presents the predicted
spectral efficiency vector and \(\mathbf{e}_i\) is the prediction error. This
error represents the mismatch between the predicted and realized spectral
efficiency due to channel variations, mobility, fading, and prediction
inaccuracies. We model the prediction error as a Gaussian random vector. In the reference
case, we assume an unbiased prediction process, i.e.,
$\mathbf{e}_i
\sim
\mathcal{N}
\left(
\mathbf{0},
\boldsymbol{\Sigma}_i
\right),$
where \(\boldsymbol{\Sigma}_i\) is the covariance matrix of the prediction error. The Gaussian assumption is adopted because it leads to a tractable
deterministic reformulation of the chance constraint through the inverse
standard normal cumulative distribution function. Hence, the realized spectral efficiency vector is distributed as
$\boldsymbol{\Lambda}_i
\sim
\mathcal{N}
\left(
\overline{\boldsymbol{\Lambda}}_i,
\boldsymbol{\Sigma}_i
\right).$

The SLA requirement imposes that the demand of each cell is satisfied over the
complete BH cycle with probability at least \(1-\epsilon_i\), where
\(0<\epsilon_i<1\) is the maximum allowed SLA violation probability for cell
\(i\in\mathcal{I}\). Therefore, the requirement is expressed as
\begin{equation}
\Pr\left[
T_{\mathrm{slot}}B_{\mathrm{tot}}
\mathbf{b}_i^{\mathsf{T}}
\boldsymbol{\Lambda}_i
\ge D_i
\right]
\ge 1-\epsilon_i,
\qquad \forall i\in\mathcal{I}.
\label{eq:sla_chance_constraint_fixed}
\end{equation}
Because \(\mathbf{b}_i^{\mathsf{T}}\boldsymbol{\Lambda}_i\) is a Gaussian
random variable with mean
\(\mathbf{b}_i^{\mathsf{T}}\overline{\boldsymbol{\Lambda}}_i\) and variance
\(\mathbf{b}_i^{\mathsf{T}}\boldsymbol{\Sigma}_i\mathbf{b}_i\), its standard
deviation is:
\begin{equation}
\sqrt{
\mathbf{b}_i^{\mathsf{T}}
\boldsymbol{\Sigma}_i
\mathbf{b}_i
}
=
\left\|
\boldsymbol{\Sigma}_i^{1/2}
\mathbf{b}_i
\right\|_2.
\label{eq:sla_std_fixed}
\end{equation}
This norm representation highlights the second-order cone structure of the
deterministic SLA constraint. Let \(\kappa_i\) represent
the reliability factor associated with the SLA level \(1-\epsilon_i\) as
\begin{equation}
\kappa_i = \Phi^{-1}(1-\epsilon_i),
\label{eq:kappa_fixed}
\end{equation}
where \(\Phi^{-1}(\cdot)\) is the inverse cumulative distribution function
of the standard normal distribution. Using
this reliability factor, the chance constraint can be rewritten
deterministically as
\begin{equation}
T_{\mathrm{slot}}B_{\mathrm{tot}}
\left(
\mathbf{b}_i^{\mathsf{T}}
\overline{\boldsymbol{\Lambda}}_i
-
\kappa_i
\left\|
\boldsymbol{\Sigma}_i^{1/2}
\mathbf{b}_i
\right\|_2
\right)
\ge D_i,
\qquad \forall i.
\label{eq:sla_deterministic_fixed}
\end{equation}
Therefore, the SLA constraint can be interpreted as a demand satisfaction
condition with a reliability margin against spectral efficiency uncertainty.
More precisely, the expected delivered traffic reduced by the
reliability-dependent uncertainty margin must remain larger than the required
demand \(D_i\). This margin increases when the spectral efficiency prediction
is less accurate or when a higher reliability level is required.
Based on the deterministic reformulation in
\eqref{eq:sla_deterministic_fixed}, the SLA-aware BH optimization problem is
formulated by considering the expected total delivered traffic as the objective:
\begin{subequations}
\label{eq:P2_fixed_sla}
\begin{align}
\max_{\substack{\mathbf{x}^t,\mathbf{b}^t\\ t\in\mathcal{T}}}\quad
& \sum_{t\in\mathcal{T}}\sum_{i\in\mathcal{I}}
T_{\mathrm{slot}}
B_{\mathrm{tot}} b_i^t
\overline{\Lambda}_i^t
\label{eq:P2_fixed_obj}
\\
\text{s.t.}\quad
& T_{\mathrm{slot}}B_{\mathrm{tot}}
\left(
\mathbf{b}_i^{\mathsf{T}}
\overline{\boldsymbol{\Lambda}}_i -
\kappa_i
\left\|
\boldsymbol{\Sigma}_i^{1/2}
\mathbf{b}_i
\right\|_2
\right)
\ge D_i,
\qquad \forall i\in\mathcal{I},
\label{eq:P2_fixed_sla_constraint}
\\
& \mathbf{1}^{\mathsf{T}}\mathbf{x}^t \le K_{\max},
\qquad \forall t\in\mathcal{T},
\label{eq:P2_fixed_active_beams}
\\
& \mathbf{1}^{\mathsf{T}}\mathbf{b}^t \le 1,
\qquad \forall t\in\mathcal{T},
\label{eq:P2_fixed_bandwidth_budget}
\\
& \mathbf{0} \le \mathbf{b}^t \le b_{\max}\mathbf{x}^t,
\qquad \forall t\in\mathcal{T},
\label{eq:P2_fixed_bandwidth_activation}
\\
& \mathbf{x}^t\in\{0,1\}^{N},
\qquad \forall t\in\mathcal{T}.
\label{eq:P2_fixed_binary}
\end{align}
\end{subequations}
%The objective function in \eqref{eq:P2_fixed_obj} maximizes the expected
%total delivered traffic. It is written using the predicted spectral efficiency \(\overline{\Lambda}_i^t\), since it represents the expected value of the random spectral efficiency. 
Constraint
\eqref{eq:P2_fixed_sla_constraint} guarantees that the demand of each cell is
satisfied over the complete BH cycle with probability at least~\(1-\epsilon_i\). 
%Constraints
%\eqref{eq:P2_fixed_active_beams}--\eqref{eq:P2_fixed_binary} define the payload activation and bandwidth allocation limits.
\section{Simulation Results}
\label{sect5}
\begin{table}[t!]
\vspace*{0.04in}
\centering
\caption{Main Simulation Parameters}
\label{tab:simulation_parameters}
\begin{tabular}{lll}
\textbf{Parameter}
& \textbf{Value} \\
\hline
$N$
& $48$ \\
$T$
& $64$ \\
\(T_{\mathrm{slot}}\) & 2 ms \\
$B_{\mathrm{tot}}$
& $500$ MHz \\
$P_{\mathrm{tot}}$ & $1000$ W \\
$K_{\max}$
& $32$ \\
$b_{\max}$
& $0.6$ \\
$\epsilon_i$
& $0.05$ \\
$1-\epsilon_i$
& $95\%$ \\
Uncertainty level & 40\% \\
BH cycle duration & 128 ms \\
 Packet size
& $1500$ bytes \\
\hline
\end{tabular}
\end{table}
In this section, we evaluate the performance of the chance-constrained
optimization model in \eqref{eq:P2_fixed_sla}. The proposed method is compared
with the baseline BH model in \eqref{eq:P1_fixed} and with an adapted OCDR
benchmark based on \cite{9146871}. To ensure a fair comparison, the OCDR model
is extended by including bandwidth allocation as a decision variable, and it
maximizes the minimum offered-capacity-to-demand ratio among the cells. The
main simulation parameters are summarized in Table~\ref{tab:simulation_parameters}.
The predicted spectral efficiency matrix
\(\overline{\boldsymbol{\Lambda}}\) is generated for all cells and time slots
from a shadowed Rician channel model with moderate line-of-sight shadowing.
The prediction error is modelled as a zero-mean Gaussian random variable whose
standard deviation is proportional to the predicted spectral efficiency, i.e.,
\(\sigma_i^t=\delta\overline{\Lambda}_i^t\), where \(\delta\) is the relative
uncertainty level. Assuming independent prediction errors across time slots,
the covariance matrix of the cell \(i\) is given by
$\boldsymbol{\Sigma}_i =
\mathrm{diag}\left((\sigma_i^1)^2,\ldots,(\sigma_i^T)^2\right).$
After solving each optimization problem, the obtained BH pattern and bandwidth
allocation are evaluated through Monte Carlo simulations. In each realization,
the actual spectral efficiency is generated as
\(\Lambda_i^t=\overline{\Lambda}_i^t+e_i^t\), and the delivered traffic is
computed accordingly. The empirical SLA violation probability, SLA
satisfaction, unmet demand ratio, and fairness are then evaluated. The same
Monte Carlo realizations are used for all compared methods, and we report \(95\%\) confidence intervals where applicable. All optimization problems were solved using Gurobi with a \(300\)s time limit, one computational thread, and a fixed solver seed. The achieved optimality gaps were \(0\%\), \(1.29\%\), and \(0.07\%\) for the deterministic baseline, OCDR-BH, and the proposed SLA-aware method, respectively.
\begin{figure}[t!]
    \centering
   \includegraphics[width=0.45\textwidth]{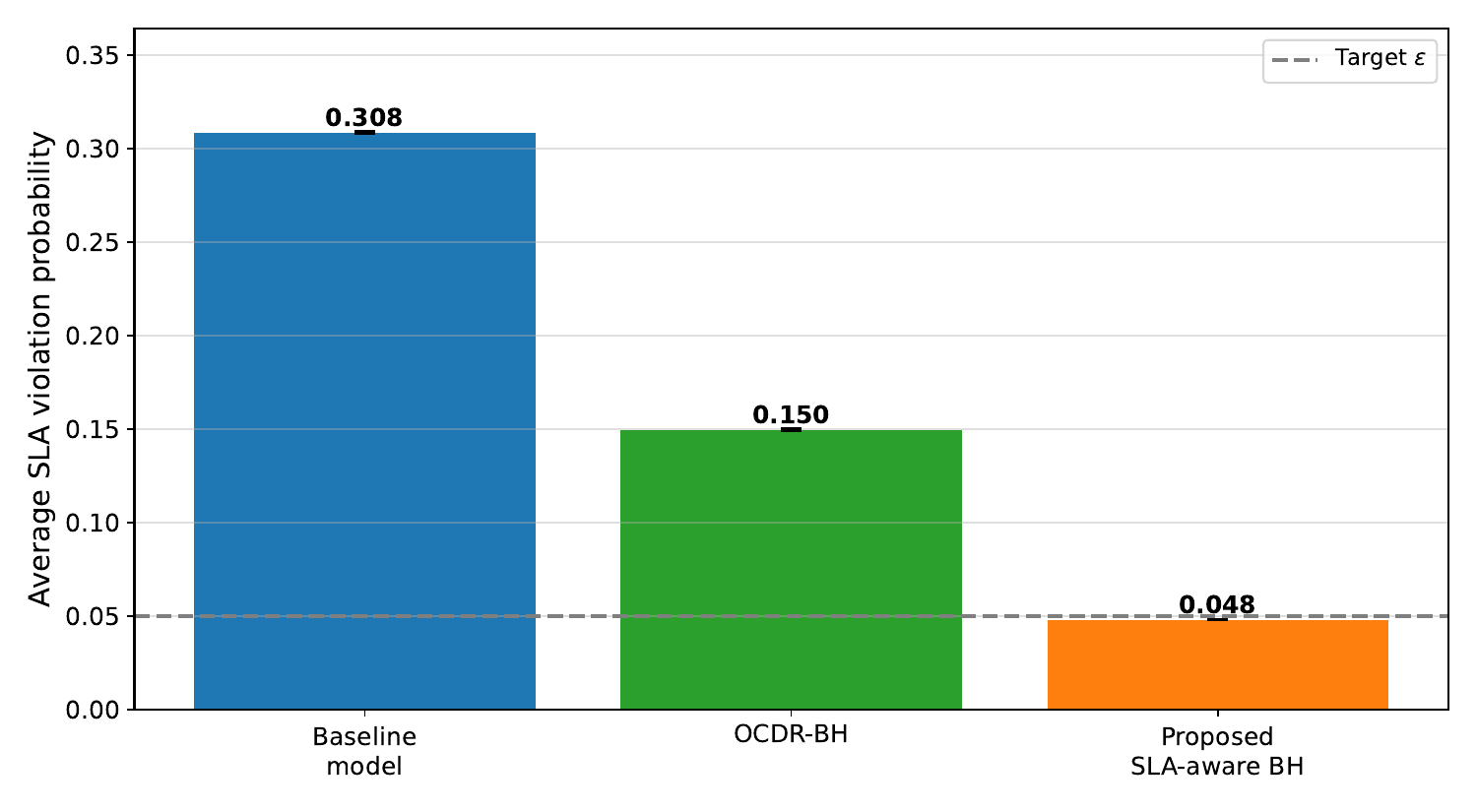}
          \caption{Average SLA violation of the compared BH schemes under the reference scenario.}
        \label{fig1}
\end{figure}

\begin{figure}[t!]
   \includegraphics[width=0.45\textwidth]{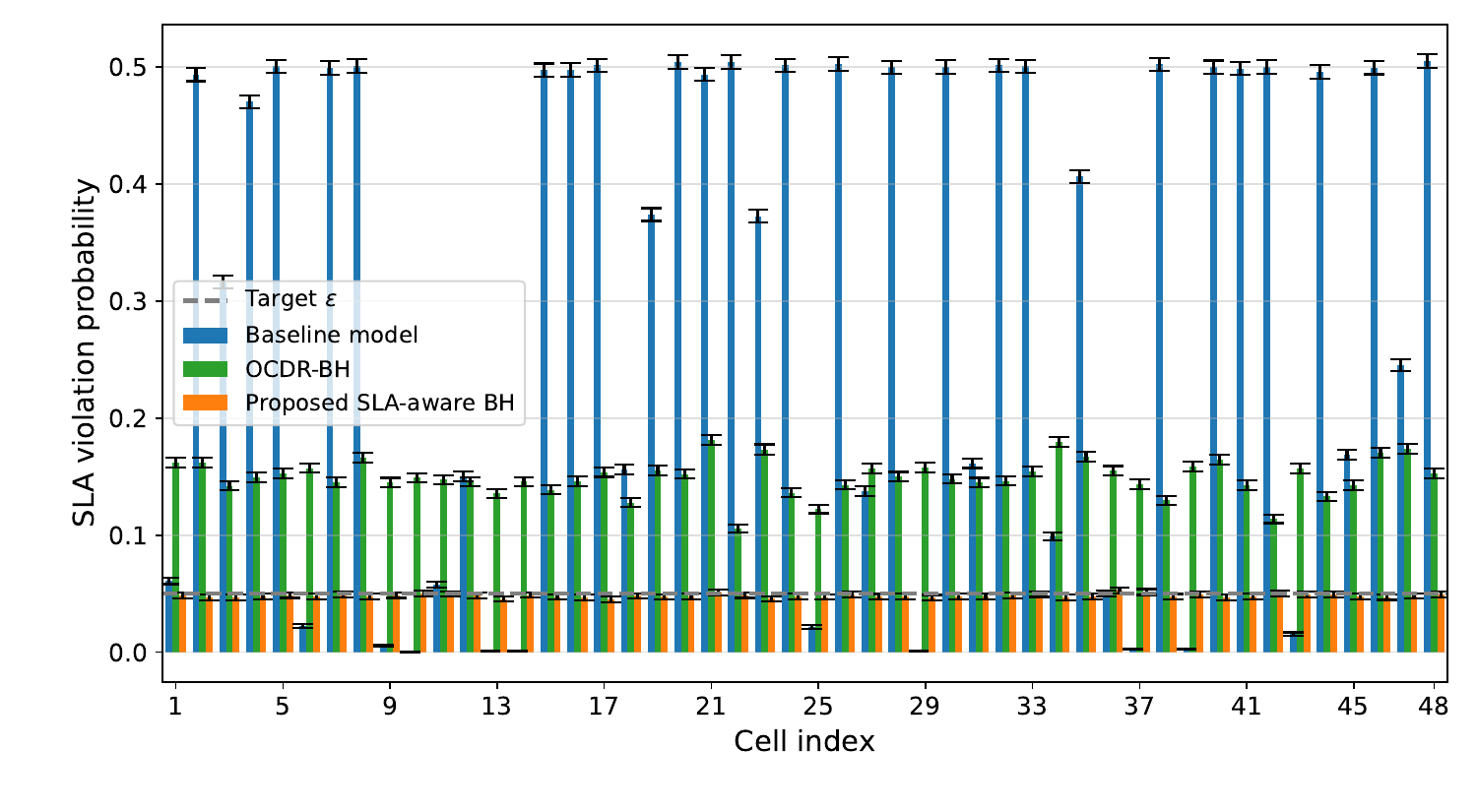}
       \caption{Per-cell SLA violation probability of the compared BH schemes under the reference scenario.}
        \label{fig6}
\end{figure}

\begin{figure}[t!]
    \centering
   \includegraphics[width=0.44\textwidth]{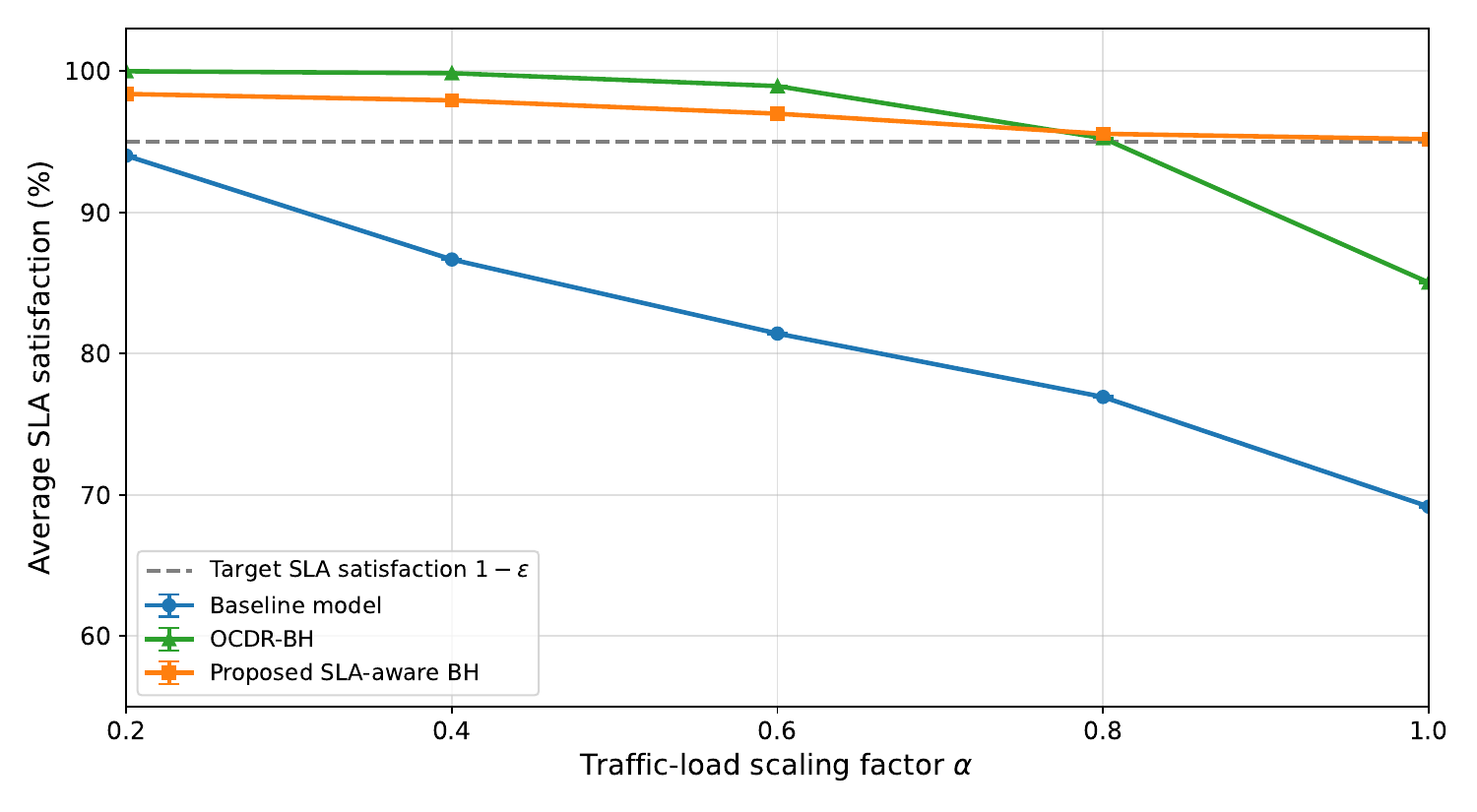}
          \caption{Average reliability versus the traffic-load scaling factor.}
        \label{fig2}
\end{figure}

Fig.~\ref{fig1} and \ref{fig6} compare the SLA violation probability of the three methods under the reference scenario. As shown in Fig.~\ref{fig1}, the baseline model has the highest average violation probability, equal to \(0.308\), while OCDR-BH reduces it to \(0.150\). The proposed SLA-aware BH method achieves \(0.048\), which is close to the target value \(\epsilon=0.05\). Fig.~\ref{fig2} confirms this behaviour at the cell level. The baseline model shows large violations for several cells, and OCDR-BH reduces them but remains above the target for many cells. In contrast, the proposed method keeps most cell-level violations close to the target. These results confirm that explicitly accounting for spectral efficiency uncertainty in the optimization is crucial for maintaining SLA violation probabilities within the target reliability level.

Fig.~\ref{fig2} shows the average SLA satisfaction for different
traffic-load scaling factors. The traffic-load factor \(\alpha\) scales the
reference demand according to \(D_i(\alpha)=\alpha D_i\). As
shown, the SLA satisfaction of the baseline model decreases significantly when
the traffic load increases. The OCDR-BH method provides better reliability,
especially at low and medium traffic loads, because it balances the offered
capacity with respect to the cell demands. However, its performance decreases
at higher traffic loads since it does not explicitly account for
spectral efficiency uncertainty. In contrast, the proposed SLA-aware BH method
maintains an SLA satisfaction close to the target value over the considered
traffic-load range.

% shows the average reliability for different traffic-load scaling factors. The SLA satisfaction of the baseline model decreases significantly when the traffic load increases. The OCDR method provides better reliability, especially at low and medium traffic loads, but its performance decreases at higher loads. In contrast, the proposed method maintains an SLA satisfaction close to the target value over the considered traffic-load range.
\begin{figure}[t!]
    \centering
   \includegraphics[width=0.44\textwidth]{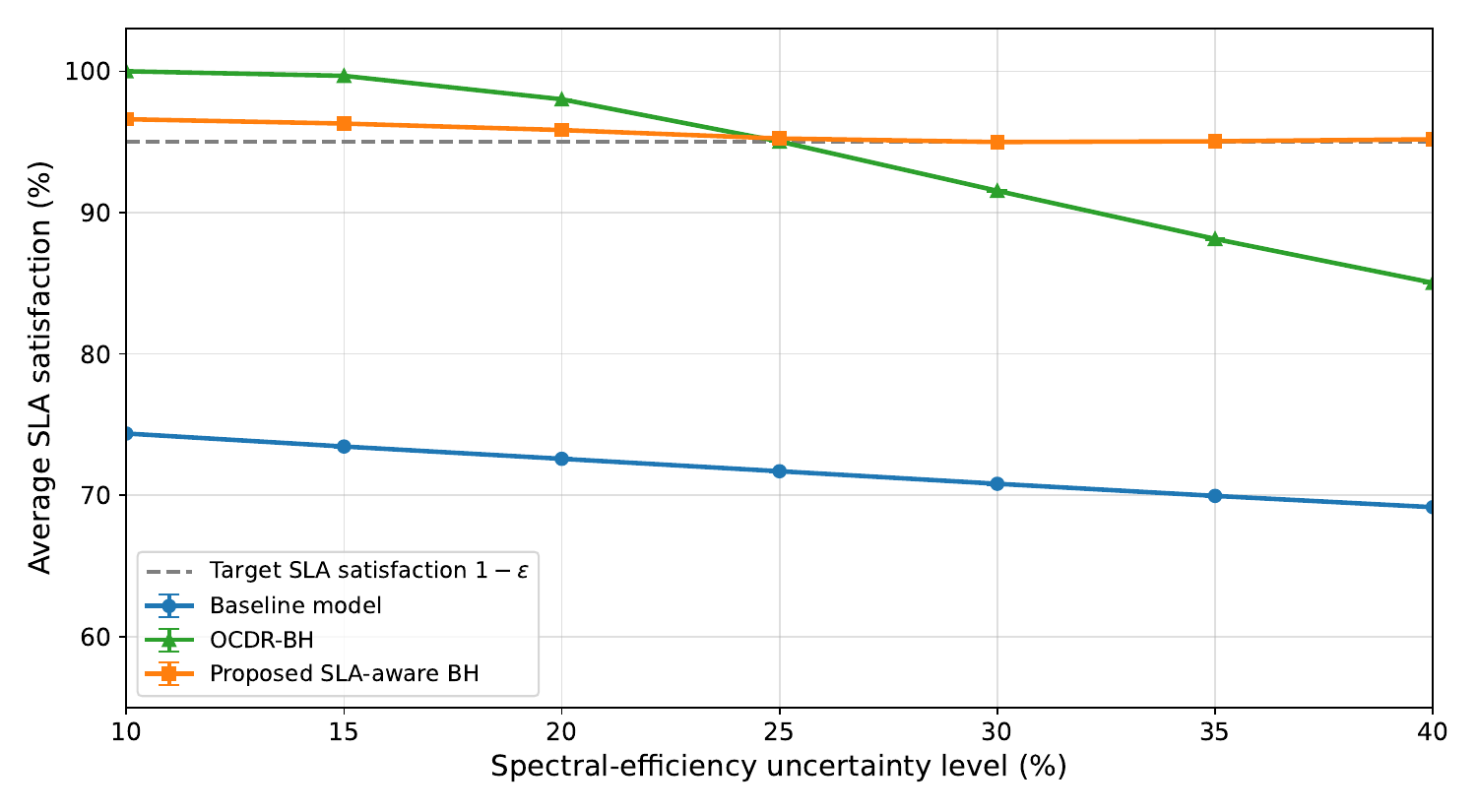}
          \caption{Average reliability versus the spectral efficiency uncertainty level.}
        \label{fig3}
\end{figure}
\begin{figure}[t!]
    \centering
   \includegraphics[width=0.44\textwidth]{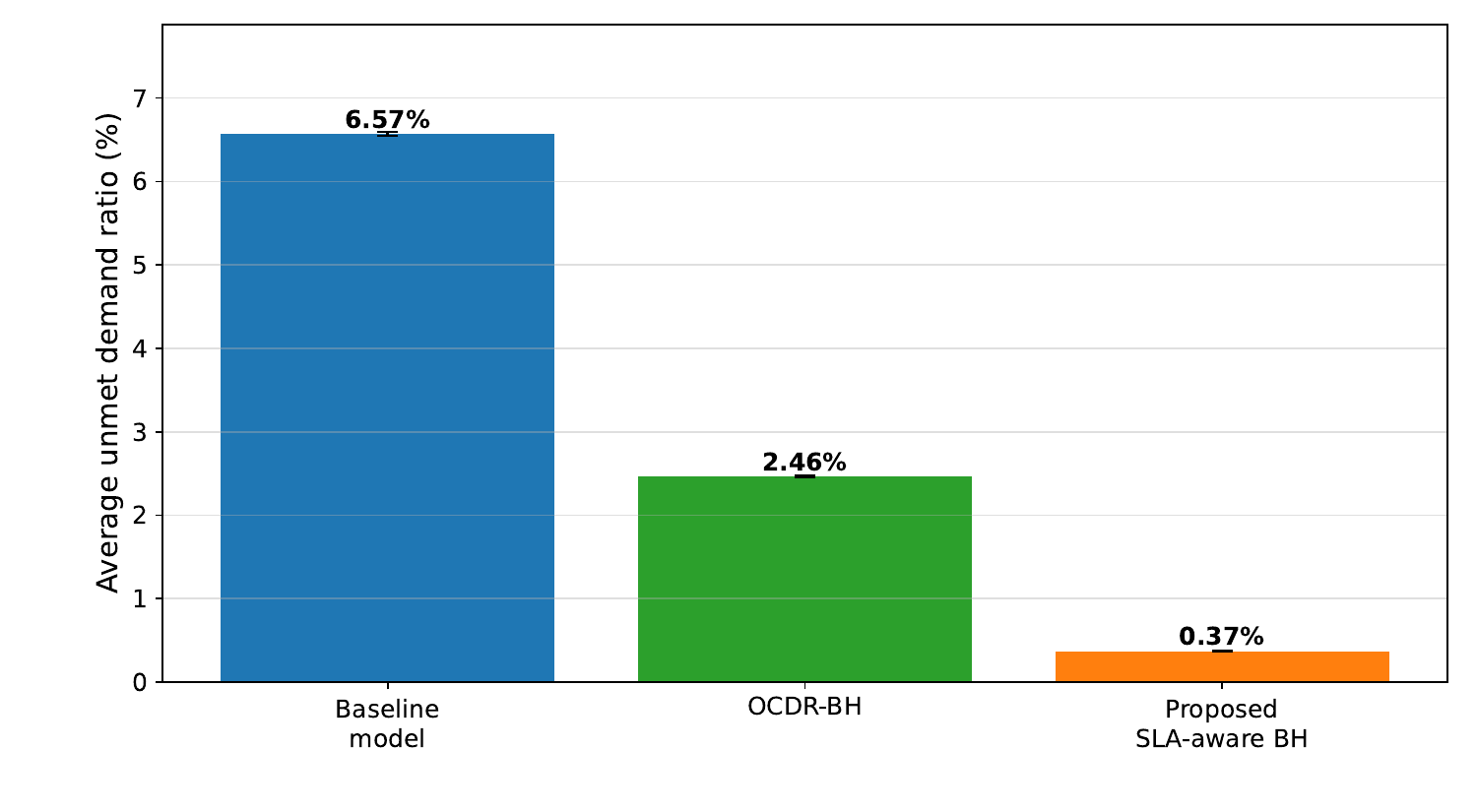}
          \caption{Average unmet demand ratio of the compared BH schemes.}
        \label{fig4}
\end{figure}
Fig.~\ref{fig3} illustrates the impact of the relative
spectral efficiency uncertainty level \(\delta\) on the average SLA
satisfaction. As shown, the reliability of the baseline and OCDR methods decreases with increasing uncertainty because these methods do not explicitly account for channel uncertainty. However, the proposed method maintains an SLA satisfaction close to the target value $95\%$ for all the considered uncertainty levels.

%\begin{figure}[t!]
 %   \centering
  % \includegraphics[width=0.48\textwidth]{cdf_per_cell_violation_probability.pdf}
   %       \caption{Empirical CDF of the per-cell violation probability.}
    %    \label{fig4}
%\end{figure}

Fig.~\ref{fig4} compares the average unmet demand ratio of the three
methods. The baseline model has the highest unmet demand ratio, equal to
\(6.57\%\), while the adapted OCDR method reduces it to \(2.46\%\). The proposed chance-constrained BH method achieves the lowest unmet demand ratio at \(0.37\%\). This confirms that the proposed method improves service reliability by reducing unmet demand under spectral efficiency uncertainty.

 \begin{figure}[t!]
    \centering
   \includegraphics[width=0.44\textwidth]{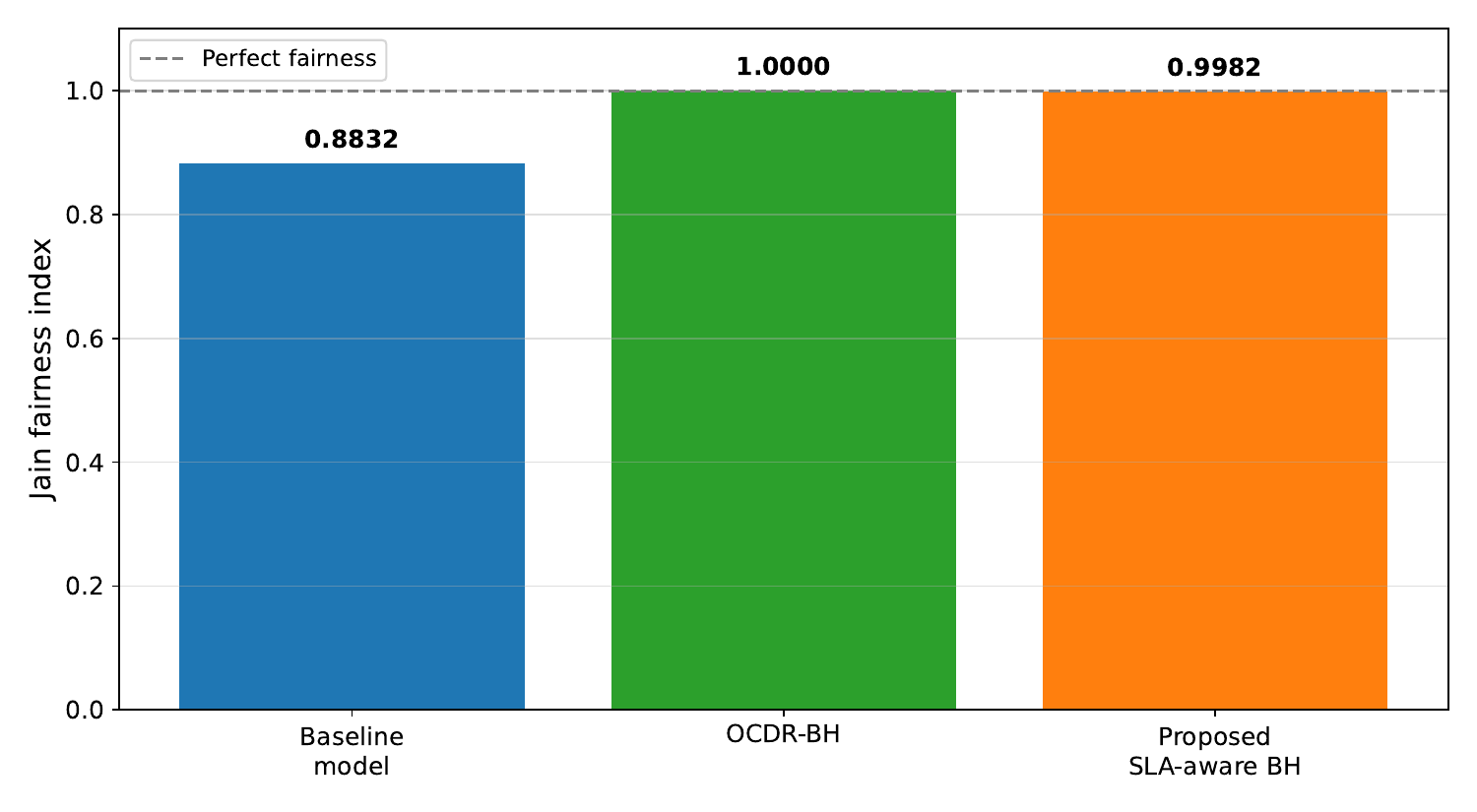}
          \caption{Jain fairness index of the compared BH schemes.}
        \label{fig5}
\end{figure}

Fig.~\ref{fig5} compares the Jain fairness index of the three methods. The baseline BH scheme achieves a fairness index of $0.8832$, which indicates an unequal distribution of the delivered traffic among the cells. Both the OCDR and chance-constrained methods achieve fairness indices close to 1, indicating that they provide a more balanced service across cells.

\section{Conclusion}
\label{sec:conclusion}

In this paper, we proposed an SLA-aware chance-constrained optimization framework for joint beam scheduling and bandwidth allocation in multibeam satellite systems under spectral efficiency uncertainty. The proposed method maximizes the expected delivered traffic while ensuring that the demand of each cell is satisfied with a predefined reliability level. We compared our method with a baseline and OCDR BH models. The numerical results showed that the proposed chance-constrained method maintains a reliability close to the target value of \(95\%\). It also achieves a Jain fairness index close to one, indicating a fair resource distribution among the cells. These results show that accounting for spectral efficiency uncertainty is important for reliable BH resource allocation. In future work, we will consider dynamic beam-to-cell association and joint power allocation.
\section*{Acknowledgement}
%\balance
This
work was supported in part by MDA; in part by the Consortium de Recherche
et d’innovation en Aérospatiale au Québec (CRIAQ); in part by MITACS;
and in part by the Natural Sciences and Engineering Research Council
of Canada (NSERC).
\bibliographystyle{IEEEtran}
\bibliography{reference}
\end{document}